\documentclass{nature}
\usepackage{graphicx,threeparttable,epsfig,multirow,booktabs,rotating}
\usepackage[none]{hyphenat}

\newcounter{firstbib}

\title{Relativistic baryonic jets from an ultraluminous supersoft X-ray source}

\author{Ji-Feng Liu$^{1}$,
Yu Bai$^{1}$,
Song Wang$^{1}$,
Stephen Justham$^{1}$,
You-Jun Lu$^{1}$,
Wei-Min Gu$^{2}$,
Qing-Zhong Liu$^{3}$,
Rosanne Di Stefano$^{4}$,
Jin-Cheng Guo$^{1}$,
Antonio Cabrera-Lavers$^{5,6}$,
Pedro \'{A}lvarez$^{5,6}$,
Yi Cao$^{7}$,
Shri Kulkarni$^{7}$,
}

\def\lesssim{\mathrel{\hbox{\rlap{\hbox{\lower4pt\hbox{$\sim$}}}\hbox{$<$}}}}
\def\gtrsim{\mathrel{\hbox{\rlap{\hbox{\lower4pt\hbox{$\sim$}}}\hbox{$>$}}}}
\def\ggg{\mathrel{\hbox{\rlap{\hbox{\lower4pt\hbox{$\sim$}}}\hbox{$>$}}}}

\begin{document}

\maketitle

\begin{affiliations}
\item Key Laboratory of Optical Astronomy, National Astronomical Observatories, Chinese Academy of Sciences, 20A Datun Rd, Chaoyang District, Beijing, China 100012
\item Department of Astronomy, Xiamen University, Xiamen, Fujian Province, China 361005
\item Key Laboratory of Dark Matter and Space Astronomy, Purple Mountain Observatory, Chinese Academy of Sciences, Nanjing, Jiangsu Province, China 210008
\item Harvard-Smithsonian Center for Astrophysics, Cambridge, MA, USA 02138
\item Instituto de Astrof\'{i}sica de Canarias, c/V\'{i}a L\'{a}ctea s/n, E-38200 La Laguna, Tenerife, Spain.
\item Departamento de Astrof\'{i}sica, Universidad de la Laguna, E-38205 La Laguna, Tenerife, Spain.
\item Department of Astronomy, Caltech, Pasadena, CA, USA 91125
\end{affiliations}

\begin{abstract}


The formation of relativistic jets by an accreting compact object is one of the
fundamental mysteries of astrophysics.  While the theory is poorly understood,
observations of relativistic jets from systems known as
microquasars\cite{Mirabel98,Paredes03} have led to a well-established
phenomenology\cite{Fender04,Migliari06}.
Relativistic jets are not expected
from sources with soft or supersoft X-ray spectra, although two such systems
are known to produce relatively low-velocity bipolar
outflows\cite{Southwell96,Becker98}.
Here we report optical spectra of an ultraluminous supersoft X-ray source
(ULS\cite{DiStefano03,Swartz02}) in the nearby galaxy M81 (M81
ULS-1\cite{Liu08a,Liu08b}) showing blueshifted broad H$_\alpha$ emission lines,
characteristic of baryonic jets with relativistic speeds.
The time variable jets have projected velocities $\sim$17 per cent of the speed
of light, and seem similar to those in the prototype microquasar SS
433\cite{Margon84,Blundell07}.
Such relativistic jets are not expected to be launched from white
dwarfs\cite{Livio01}, but an origin from a black hole or neutron star in M81
ULS-1 is hard to reconcile with its constant soft X-rays\cite{Liu08b}.
The completely unexpected presence of relativistic jets in a ULS challenges the
canonical theories for jet formation\cite{Fender04,Migliari06}, but may
possibly be explained by a long speculated super-critically accreting black
hole with optically thick
outflows\cite{Paczynsky80,Abramowicz88,Jiang14,Sadowsky15,Ohsuga05,King03,Shen15}.

\end{abstract}

Initial spectroscopic observations of M81 ULS-1 made at the Keck Observatory in
2010 found\cite{Bai15}  broad Balmer emission lines (as wide as 400 km/s) on
top of a power-law like blue continuum.
A very broad emission line (as wide as 30\AA, corresponding to 2000 km/s) was
detected around 5532/5543\AA\ in both observations, but was not identified with
any known spectral lines.
We followed up with new spectra obtained at the {\it Gran Telescopio Canarias}
(GTC) in 2015, which again showed the Balmer emission lines and the blue
continuum, but the previously-unidentified broad emission line was now at a
significantly changed wavelength of 5648\AA\ (Figure~1).
This change in observer-frame wavelength immediately suggests that the
previously unidentified emission line is a blue-shifted H$_\alpha$ emission
line emitted by an approaching baryonic relativistic jet, at projected
velocities of $-$0.17$c$.
More subsequent follow-up spectra reveal ongoing changes in the projected
velocity of the blue-shifted jet, for which we suggest that the best
explanation is jet precession, as observed in the prototype microquasar SS~433.

SS~433 has exhibited time-variable blueshift and redshift of optical emission
lines emitted by its precessing jets, the long-term monitoring of which has
revealed\cite{Margon84,Blundell07} a precession period of 164 days, and an
intrinsic jet velocity of 0.26$c$.
M81 ULS-1 is only the second microquasar identified through directly measuring
the blueshift of H$_\alpha$ emission lines emitted by its baryonic jets.
Other known microquasars\cite{Mirabel98,Paredes03} have mostly been identified
through direct imaging of their radio jets or by interpreting strong
non-thermal radio emission as arising from their relativistic jets with
velocities above 0.1$c$.
M81 ULS-1 has not previously been detected by radio surveys, but this is not
surprising given the great distance to M81.
Were SS~433 placed in M81, its radio flux at Earth would be about 1$\mu$Jy,
below the detection sensitivity of current radio facilities, but achievable in
the future with the Square Kilometer Array\cite{ska}.

From its X-ray properties\cite{Liu08b}, M81 ULS-1 is a truly unique jet source,
in distinct contrast to all other known microquasars\cite{Mirabel98,Paredes03}.
During all observations of M81 by the Chandra X-ray Observatory since its
launch, ULS-1 is always detected and exhibits high and low flux states with
count rates ranging from 1 to 70 photons per kilo-second.
When in high flux states, M81 ULS-1 clearly exhibits supersoft spectra with
blackbody temperatures of 65-100 eV and bolometric luminosities above $10^{39}$
erg s$^{-1}$.  Somewhat surprisingly, the low flux state of M81 ULS-1 appears
as supersoft as the high flux state, with more than 95\% of the photons below 1
keV (Figure~2).
In contrast, all known microquasars\cite{Mirabel98,Paredes03} are low-mass or
high-mass X-ray binaries, each measured or thought to contain a neutron star or
black hole, emitting abundant hard photons above 1 keV. As shown in Fig~2, only
a few percent to 35\% of their photons are below 1 keV in Chandra observations.

Luminous supersoft sources\cite{Kahabka97} (SSSs) have supersoft X-ray spectra,
and those with  luminosities below the Eddington limit for a solar-mass object
are conventionally interpreted as white dwarfs accreting at a rate of about
$\rm 10^{-7}-10^{-6}~M_\odot~yr^{-1}$, for which hydrogen fusion within the
accreted material proceeds steadily\cite{Iben82,Nomoto82}, but the presence of
relativistic jets suggests otherwise in M81 ULS-1.
The relativistic jets observed in ULS-1 are simply not expected for typical
white dwarfs\cite{Livio01}.
Indeed, while low velocity bipolar outflows of a few thousand $\rm km~s^{-1}$
are possible and have been observed in SSSs such as RX
J0513.9-6951\cite{Southwell96} and RX J0019.8+2156\cite{Becker98}, no
relativistic jets have ever been observed from SSSs other than M81 ULS-1.
These considerations suggest that the accreting object in M81 ULS-1 is not a
white dwarf, adding strong evidence to the idea\cite{distefano04,distefano10}
that SSSs, especially the ultraluminous ones, do not necessarily contain
accreting white dwarfs.

If the central engine of M81 ULS-1 is indeed a neutron star or black hole as in
all other known microquasars, established
phenomenology\cite{Fender04,Migliari06} expects steady jets to be generated in
the low-hard state, with episodic jets generated in the very high state or
during the state transitions between soft and hard states.
In the case of M81 ULS-1, the blue-shifted H$_\alpha$ emission lines emitted
from the relativistic jets were present in all six optical spectroscopic
observations in 2010 and 2015.
Standard presumptions would therefore be that M81 ULS-1 is in the low-hard or
very high states for a substantial fraction of the time, during which abundant
hard photons above 1 keV are expected as in other microquasars (Fig~2).
Contrary to the above expectations,  ULS-1 has been persistently supersoft in
all 19 Chandra observations, regardless of whether it is displaying a low or
high flux state, suggesting that its relativistic jets are not generated in the
canonical ways.
In fact, the persistently supersoft appearance of ULS-1 is not expected in any
spectral states in the standard accretion scenarios\cite{RM06,Hasinger89} for
neutron star or black hole X-ray binaries, which are known to be accreting
below the critical (i.e., Eddington) rate.

This unusual combination of relativistic jets and persistently supersoft X-ray
spectra is completely unexpected, posing challenges to our conventional
understanding of jet formation\cite{Fender04,Migliari06}.
One possible explanation for M81 ULS-1 can be a long
speculated\cite{Paczynsky80,Abramowicz88} super-critically accreting black hole
with optically thick outflows.
Recent magneto-hydrodynamic simulations of such systems, although still under
development and in heated debate\cite{Jiang14,Sadowsky15}, can generate
super-Eddington luminosities, and necessarily\cite{Gu15} generate disk winds
and funnels along the rotation axis, from which radiation pressure will drive
baryon-loaded relativistic jets with velocities up to $0.3c$ regardless the
black hole spin\cite{Sadowsky15}.
M81 ULS-1 qualitatively matches both predictions of high luminosities and
baryon-loaded relativistic jets, and its supersoft X-ray spectra can
potentially be expected from optically thick outflows for suitable outflow
geometry, wind velocities and outflow mass rates\cite{Ohsuga05,King03,Shen15}.
Thus, ULS-1 may potentially be a vivid manifestation of recent predictions for
super-critical accretion onto black holes,  and so reveal the nature of extreme
accretion in extreme conditions.

\begin{addendum}

 \item[Acknowledgements]  We thank Drs Katherine Blundell, Ramesh Narayan,
	 Zhiyuan Li, Tinggui Wang, Feng Yuan, Xuan Fang, Jimmy Irwin, Tom
	 Maccarone and Doug Swartz for helpful discussions. Authors acknowledge
	 supports from the Chinese Academy of Sciences through Grant No.
	 XDB09000000, from 973 Program 2014CB845705, and from National Science
	 Foundation of China through grants NSFC-11333004/11425313. This work
	 is partly based on observations made with the Gran Telescopio Canarias
	 (GTC), installed in the Spanish Observatorio del Roque de los
	 Muchachos of the Instituto de Astrof¨ªsica de Canarias, in the island
	 of La Palma.  Part of the data was obtained at the W.M. Keck
	 Observatory, which is operated as a scientific partnership among the
	 California Institute of Technology, the University of California and
	 the National Aeronautics and Space Administration. The Observatory was
	 made possible by the generous financial support of the W.M. Keck
	 Foundation.

 \item[Author contributions] J.-F.L. proposed the observations. J.-F.L., Y.B.
	 and S.W. reduced the optical and X-ray data and carried out the
	 analysis.  J.-F.L., S.J., Y.-J.L. and R.D.S. discussed the results and drafted
	 the manuscript. All authors commented on and helped improving the
	 manuscript.


 \item[Author information] Reprints and permissions information is available at
	 npg.nature.com/reprints.  The authors declare that they have no
	 competing financial interests.  Readers are welcome to comment on the
	 online version of the paper.  Correspondence and requests for
	 materials should be addressed to J.F. Liu  (email: jfliu@nao.cas.cn).

\end{addendum}

\begin{figure}
\includegraphics[width=6.0in,height=8.0in]{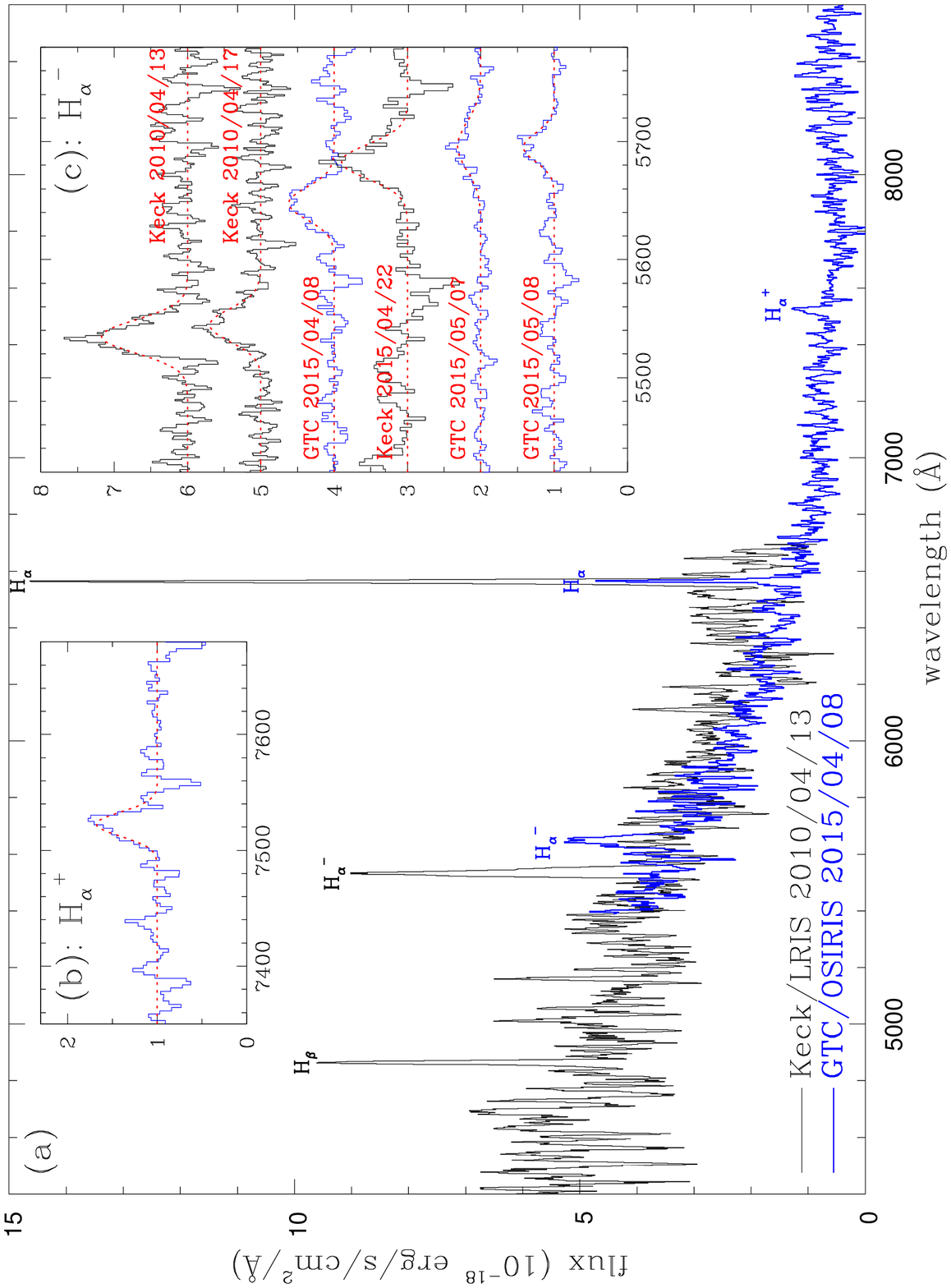}
\end{figure}

\begin{figure}
\includegraphics[width=5.4in,height=5.4in]{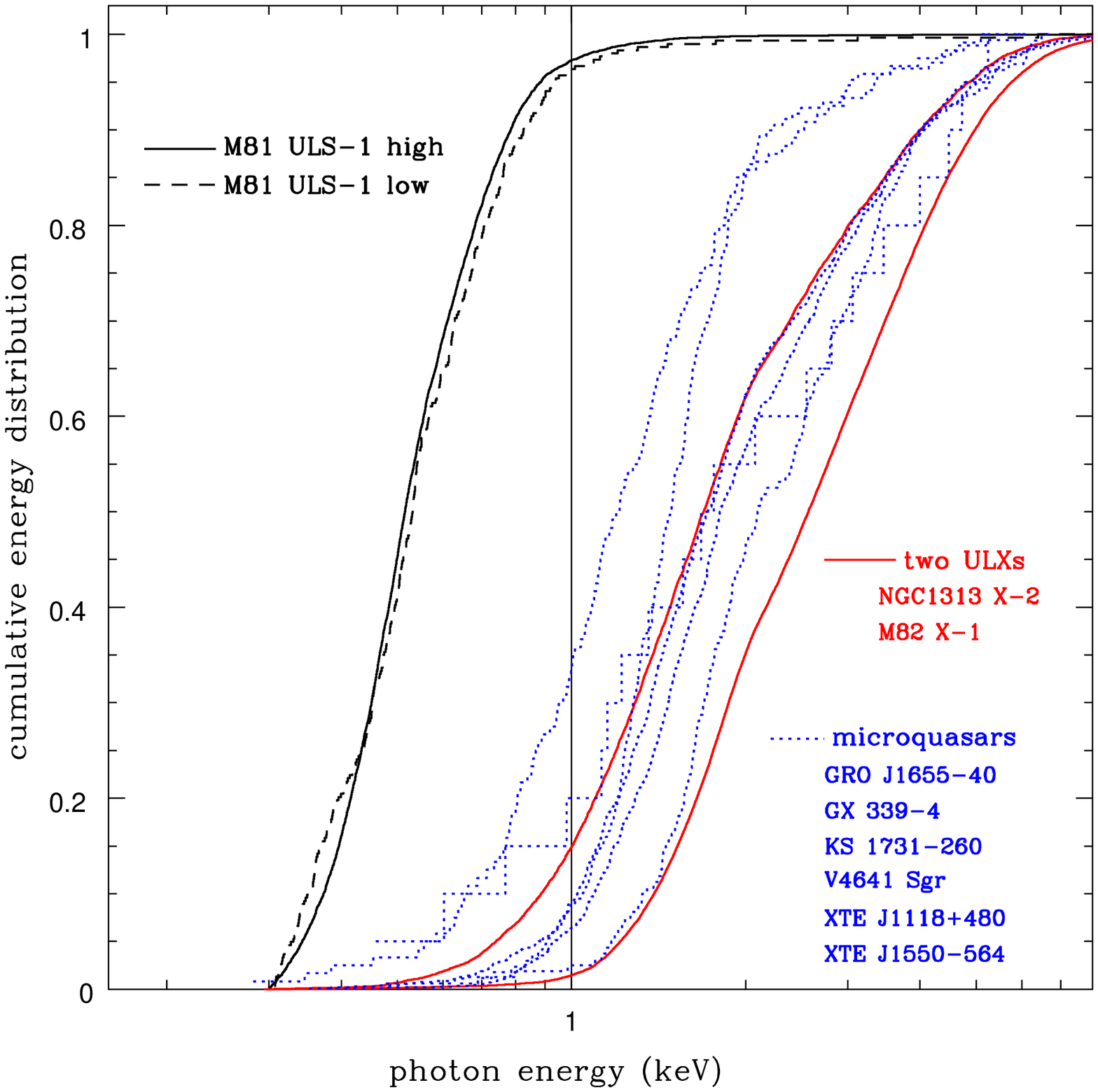}
\end{figure}

\noindent{\bf Figure legends}\\
{\bf Figure 1. Keck and GTC spectra for the optical counterpart of M81 ULS-1.}
(a) The Keck/LRIS spectrum taken on 2010/04/13 (blue channel; black) and the
GTC/OSIRIS spectrum taken on 2015/04/08 (red) for M81 ULS-1.  Labeled over the
lines are the broad Balmer lines (H$_\alpha$ and H$_\beta$), the very broad
blue-shifted H$_\alpha^-$ at 5530\AA\ and 5648\AA.  The power-law like
continuum and the broad Balmer lines are characteristic of an accretion disk
around a compact object, which confirms the physical association between the
X-ray source and its optical counterpart.  (b) The blue-shifted H$_\alpha^-$
from six Keck and GTC observations with time-variable observer-frame central
wavelengths. The intensities also change with time in proportion to the
intensities of the stationary H$_\alpha$ emission line from the accretion disk,
suggesting a link between the accretion and the jet. See the Methods for
details.  Both the spectra and the fits are normalized by the underlying
continuum, and are shifted vertically for clarity. \\
{\bf Figure 2. Cumulative photon energy distributions for M81 ULS-1.} Also
plotted for comparison are two typical ultraluminous X-ray sources in nearby
galaxies (ULXs; M82 X-1 and NGC1313 X-2), and six Galactic microquasars
observed with ACIS aboard the Chandra X-ray Observatory. Most photons
concentrate at energies where the cumulative distribution rises fastest. For
M81 ULS-1, there are more than 95\% of the photons below 1 keV both in the low
state (black solid) and the high state (red solid).  There are less than 15\%
of the photons below 1 keV for both ULXs (red solid). For all six Galactic
microquasars (blue dotted), there are only a few percent to less than 35\% of
the photons below 1 keV. Since the response matrix is not well calibrated below
0.3 keV for Chandra/ACIS, only photons above 0.3 keV are used for the above
sources.\\


\begin{methods}

%

\subsection{GTC/OSIRIS and Keck/LRIS data reduction}

Initial optical spectroscopic observations of M81 ULS-1 were carried out with
Keck/LRIS on 2010 April 13 and 17, which revealed broad Balmer emission lines as
from an accretion disk\cite{Bai15}.
The emission line of the blue-shifted H$_{\alpha}$ (H$_{\alpha}^-$) is shown at
5530 {\AA} in the spectra, with a shift in the line centre of 10 {\AA} $\pm$ 2 {\AA}
between those two observations (i.e., two epochs separated by four nights).

M81 ULS-1 was later observed using GTC/OSIRIS on 2015 April 8, May 7 and 8,
masked with the 0.6$''$ slit followed by the R1000R grating, which yields a
resolution of about 7 {\AA}.  The spectra were reduced in a standard way with
IRAF.  After bias subtraction and flat correction, the dispersion
correction was carried out based on the line lists given in the OSIRIS manual
(see http://www.gtc.iac.es/instruments/osiris/).  Raw spectra were then
extracted with an aperture size of 1$''$, and a standard star taken at each
night was used to make the flux calibration.

On 2015 April 22, another observation of M81 ULS-1 was carried out using
Keck/LRIS with the 1.0$''$ slit.  The light was split with a beam dichroic of
6800 {\AA} to the blue and red sides followed by the 300/5000 and 400/8500
gratings, which yields a resolution of $\sim$ 8 {\AA}. The spectrum was reduced
with the IDL pipeline designed for Keck.

Extended Data Table~1 lists the basic information of the observations from 2010
to 2015.
Both H$_{\alpha}$ and H$_{\alpha}^-$ emission lines are detected in all the
spectra, and their line properties are calculated from Gaussian line profile
fitting. Extended Data Table~2 lists the central wavelength, the full width at
half-maximum (FWHM) and the equivalent width (E.W.) for each fitted emission
line. The observed H$_{\alpha}^-$ central wavelengths $\lambda_-$ correspond to
projected velocities $v_r$ from $-$0.17c to $-$0.14c in these observations
given $\lambda_-/\lambda_0 = \sqrt{1+v_r/c \over 1-v_r/c}$.

\subsection{Properties of the emission lines}

In the case of SS433, the E.W.s of H$_{\alpha}$ are tightly correlated with the
phases of the precession, and those of the H$_{\alpha}^-$ emission follow a
similar trend but with a phase delay\cite{Vittone83}. We use the power of the
emission lines, calculated from the area of the Gaussian fitting, as
representative of the emission intensity, since the observed continuum from M81
ULS-1 varies significantly between observations.  Extended Data Figure~1 shows
that the power of H$_{\alpha}$ emission lines from the accretion disk is
positively correlated with that of the H$_{\alpha}^-$ emission lines,
suggesting a link between the accretion and the jet.  The variations of the
emission-lines' power are asymmetrical with smooth rises and steeper declines
around 2015 May 7 (Extended Data Figure~2), which are similar to those of
SS433\cite{Vittone83}.

The rate at which the projected H$_{\alpha}^-$ velocity changes appears slower
during the observations in 2015 compared to those in 2010. The rate of change
in 2015 is roughly 0.8 {\AA} day$^{-1}$, wheras it was 2.6 {\AA} day$^{-1}$ in
2010; if the velocity shift is due to precession this may naturally be
explained because the 2015 observations are sampling a different part of the
precession cycle. We can estimate a minimum likely precession period by
assuming that the turning point of the precession cycle occurred around the
observations of May 7 and 8, 2015 (see Extended Data Figures~2 and 3). In that
case then, after the wavelength of the emission lines reached the maximum in
May 8 (Extended Data Figure~2), the H$_{\alpha}^-$ emission likely turned back
to the short wavelength with the rate of roughly $-0.8$ {\AA} day$^{-1}$,
indicating that the half precession period must be longer than 30 days. There
is a 115 {\AA} gap between the observations in 2010 April 13 and 2015 April 08
(Extended Data Figure~3), and if we assume that the maximum rate at which the
wavelength decreases is 2.6 {\AA} day$^{-1}$, the line H$_{\alpha}^-$ needs 44
days to move by the required amount.  Therefore, the half precession period is
likely longer than 74 days, and lower limit of the precession period is about
148 days.  More time resolved spectra are needed in order to derive an accurate
period and further characterize the apparent precession of the jets.

\subsection{Searching for the red-shifted H$_{\alpha}$ emission lines}

Given the blueshifted H$_{\alpha}^-$ emission lines from the approaching jets,
the redshifted H$_{\alpha}$ emission lines (H$_{\alpha}^+$) are expected from
the receding jets, albeit with much lower intensities due to Doppler boosting
effects.
Assuming symmetrical and steady jets, the boosting factors for the lines
emitted from the approaching and the receding jets are given by $ \mathcal{D}_-
= \frac{(1-\beta^2)^{1/2}}{1-\beta \cos\theta} >1, $ and $ \mathcal{D}_+ =
\frac{(1-\beta^2)^{1/2}}{1+\beta \cos\theta} <1, $ respectively. The total flux
of a blue- or red-shifted line in the observer frame is boosted by a factor of
$D^3$. The expected central wavelengths of the two lines are $
\lambda_-=\lambda_0/\mathcal{D}_-, $ and $ \lambda_+=\lambda_0/\mathcal{D}_+$,
and the corresponding redshifts are $z_+=\lambda_+/\lambda_0-1 >0$ and
$z_-=\lambda_-/\lambda_0-1 <0$. Both $D$ and $z$ for H$_{\alpha}^-$ in all
observations are listed in Extended Data Table~3.

We have searched for the redshifted H$_{\alpha}^+$ in all observations.  A weak
emission line feature was detected at $\sim3\sigma$ around 7524\AA\ (see
Figure~1), roughly symmetrical to H$_{\alpha}^-$ at 5648\AA, in one of the GTC
exposure during the night of April 08, 2015.
If this marginal detection were the red-shifted H$_{\alpha}^+$, its boosting
factor is $\mathcal{D}_+ = \lambda_0/\lambda_+ = 0.8722$, the ratio of the
received total flux of the blue shift line to that of the the redshift line
should be $D^3_-/D^3_+\sim 2.5$, which is roughly consistent with the
observational one ($\sim 4^{+3.5}_{-1.5}$).

However, the observed wavelength is not consistent with the expected
H$_{\alpha}^+$ wavelength given the blueshifted H$_{\alpha}^-$ at 5648\AA,
i.e., $5648=6563\left[\frac{1-\beta \cos\theta}{(1-\beta^2)^{1/2}}\right]$.
Assuming the extreme case, $\theta=0^{\rm o}$, then we have $\beta= 0.1491$. If
the receding jet has the same velocity, then the expected central wavelength of
H$^+_{\alpha}$ should be 7626.4\AA, which is about 104\AA\ larger than the
detected line. If assuming $\theta=10^{\rm o}/20^{\rm o}/30^{\rm o}$, then
$\beta = 0.152/0.160/0.177$ and then the expected central wavelength of
H$^+_{\alpha}$ is 7632/7650/7688/\AA, which is about 108/126/164\AA\ larger
than the detected line. The discrepancy becomes larger for larger inclination
angles.

This casts doubt on the identification of the 7524\AA\ line feature as
H$_{\alpha}^+$, unless the jets are asymmetrical or fast changing. We may have
not detected the redshifted H$_{\alpha}^+$, but the non-detection is not
surprising given the Doppler boosting effects, and other realistic reasons. For
example, the receding jets may be blocked by the optically thick outflows if
this system is a super-critically accreting black hole system as described in
the manuscript.
No candidate H$_{\alpha}^+$ emission lines were detected at all in the May 7
and 8 GTC observations, or in the Keck spectrum.  Even if the April 8 line were
a true H$_{\alpha}^+$ emission line, this non-detection would not be
surprising, given the lower E.W.s of H$_{\alpha}^-$ on May 7 and 8, and the
relatively lower sensitivity in the red channel of LRIS.

\subsection{Analysis of {\it  \textbf{Chandra}} data}

There have been 19 {\it Chandra}/ACIS observations of the nuclear region of
M81, where ULS-1 resides.  All of these observations were derived from the {\it
Chandra} archive and analyzed uniformly with the CIAO 4.7 software tools.
Point sources were detected with WAVDETECT on the individual {\it Chandra}
images.  As listed in Extended Data Table~4,
the photon counts
were extracted from the source ellipses enclosing 95\% of the total photons as
reported by WAVDETECT, which was run with scales of 1$''$, 2$''$, 4$''$, and
8$''$ in the 0.3--8.0 keV band.

The spectra in the high states (Count Rate $>$ 10 count ks$^{-1}$) were fitted
by absorbed blackbody models, with the spectral parameters presented in
Extended Data Table~4,
all of which show that M81 ULS-1 has
been persistently supersoft in these observations.  In addition, the spectra in
high and low states were co-added into combined high- and low-state spectra,
and were also fitted in the band of 0.3-8.0 keV.
Using the fitted absorbed blackbody model, we calculated the 0.3--8.0 keV flux,
the 0.3--8.0 keV luminosity and the bolometric luminosity with the distance of
3.63 Mpc for M81\cite{Freedman1994}.


As plotted in Extended Data Figure~4, M81 ULS-1 displays a soft excess below
0.3 keV as compared to the best-fit model for 0.3-8.0 keV.  However,
considering that the response matrix for {\it Chandra} is not well calibrated
below 0.3 keV, we refrain from interpreting this soft excess.  Nonetheless, it
is clear that M81 ULS-1 has very different spectral properties from the other
known microquasars. Moreover, these uncertainties in calibration below 0.3 keV
only have the potential for the intrinsic spectral differences between M81
ULS-1 and the other known microquasars to be even larger (i.e., the energy
distribution from M81 ULS-1 may be even softer than observed).


\subsection{Code Availability}
The optical spectra were reduced with IRAF available at http://iraf.noao.edu/.
All the emission lines in Extended Data Table~2 were fitted with the curve
fitting toolbox based on Matlab (see
http://www.mathworks.com/help/curvefit/index.html).  The {\it Chandra} archive
data were analyzed with CIAO 4.7 which can be downloaded from
http://cxc.harvard.edu/ciao/download/.

\end{methods}


\begin{table}
\caption{ }
\center
\begin{tabular}{cccc}
\hline
Date        &Telescope&Exposure Time &$\Delta \lambda$\\
            &         &   (second)   &  (\AA)         \\
\hline
2010.4.13   & Keck    & 1000$\times$3& 5     \\
2010.4.17   & Keck    & 1200$\times$2& 5     \\
2015.4.08   & GTC     & 1800$\times$3& 7     \\
2015.4.22   & Keck    & 2800$\times$2& 8     \\
2015.5.07   & GTC     & 1800$\times$3& 7     \\
2015.5.08   & GTC     & 1800$\times$4& 7     \\
\hline
\end{tabular}
\end{table}

\begin{sidewaystable}
\center
\caption{ }
\begin{tabular}{clcccccc}
\hline
HJD-2450000  & \multicolumn{4}{c}{H$_{\alpha}^{-+}$}  &\multicolumn{3}{c}{H$_{\alpha}$}     \\
      \cmidrule(l){2-5} \cmidrule(l){6-8}
     & Center &    FWHM   &E.W.  & Power    &Center &    FWHM     & Power             \\
\hline
5299.83   & 5532.3 $\pm$ 1.5 ($-$0.17$c$) &33 $\pm$ 4 &  41 $\pm$ 5 & 1.55 $\pm$ 0.20 &6562.9 $\pm$ 0.2  &10.8 $\pm$ 0.5 & 1.69 $\pm$ 0.09\\
5303.77   & 5543.1 $\pm$ 1.6 ($-$0.17$c$) &33 $\pm$ 4 &  24 $\pm$ 3 & 0.94 $\pm$ 0.13 &6562.8 $\pm$ 0.2  & 9.4 $\pm$ 0.3 & 1.30 $\pm$ 0.05\\
7121.47   & 5647.5 $\pm$ 2.3 ($-$0.15$c$) &32 $\pm$ 6 &  21 $\pm$ 5 & 0.65 $\pm$ 0.15 &6564.9 $\pm$ 0.3  & 6.7 $\pm$ 0.6 & 0.31 $\pm$ 0.04\\
7134.88   & 5683.0 $\pm$ 3.0 ($-$0.14$c$) &34 $\pm$ 7 &  33 $\pm$ 9 & 1.07 $\pm$ 0.31 &6564.4 $\pm$ 0.9  & 8.9 $\pm$ 2.1 & 0.86 $\pm$ 0.27\\
7150.42   & 5696.0 $\pm$ 3.1 ($-$0.14$c$) &46 $\pm$ 8 &  16 $\pm$ 4 & 1.08 $\pm$ 0.24 &6564.1 $\pm$ 0.1  & 7.6 $\pm$ 0.3 & 1.60 $\pm$ 0.09\\
7151.46   & 5695.2 $\pm$ 2.3 ($-$0.14$c$) &29 $\pm$ 6 &  13 $\pm$ 3 & 0.57 $\pm$ 0.15 &6564.4 $\pm$ 0.2  & 7.4 $\pm$ 0.5 & 0.79 $\pm$ 0.06\\
7121.47   & 7522.1 $\pm$ 2.7 ($+$0.14$c$) &20 $\pm$ 6 &  27 $\pm$ 7 & 0.16 $\pm$ 0.05\\
\hline
\end{tabular}
\end{sidewaystable}

\begin{table}
\caption{}
\center

\begin{tabular}{lcccc} \hline
Date     & $\lambda$(H$_\alpha^-$ or H$_\alpha^+$?) & $D$ & $z$  \\ \hline
20100413 & 5532.5      & 1.1862   & -0.1570 \\
20100417 & 5543.1      & 1.1840   & -0.1554\\
20150408 & 5647.5      & 1.1621   & -0.1395 \\
20150422 & 5683.0      & 1.1548   & -0.1341 \\
20150507 & 5696.0      & 1.1522   & -0.1321 \\
20150508 & 5695.2      & 1.1523   & -0.1322 \\
20150408 & 7524.0      & 0.8722   &  0.1465 \\ \hline
\label{tab:linec}
\end{tabular}
\end{table}

\begin{sidewaystable}
\renewcommand{\arraystretch}{0.5}
\begin{center}
\caption[]{ }
\label{spectra.tab}\small
\tabcolsep=2.pt
\begin{tabular}{ccccccccccccc}
\hline
\hline
 ObsID       &Obs Date            &ExpT       & $C_{\rm NET}$      & $C_{\rm Soft}$      &Count Rate     &$kT_{\rm bb}$   &$n_{\rm H}$                &Flux                    &$L_{X}$           &$L_{\rm bol}$         &$\chi^{2}_{\nu}/dof$ &State\\
             &                    &(ks)       &                 &                 &(count ks$^{-1})$ &(eV)      &($10^{20}$ cm$^{-2}$)   &    (ergs s$^{-1}$ cm$^{-1}$) &($10^{38}$ ergs s$^{-1}$ )  &($10^{38}$ ergs s$^{-1}$ ) &  &   \\
\hline
acis390&  2000 Mar 21 &   2.4&      140&      25&       60.81$\pm$    5.74&     145$\pm$40.9&      12.5$\pm$20.0&    1.42e-13&         2.3&             6.7&      1.390/4 &high\\
acis735&  2000 May 07 &   50.7&     3679&     2141&     67.16$\pm$    1.19&     78$\pm$1.6&       9.5$\pm$1.0&    1.99e-13&         3.1&             23.1&      1.086/41  &high\\
acis5935& 2005 May 26 &   11.1&     11&       2&        1.06 $\pm$     0.43&                &                      &              &               &                           &   &low          \\
acis5936& 2005 May 28 &   11.6&     11&       2&        0.82 $\pm$     0.40&                &                      &              &               &                           &   &low          \\
acis5937& 2005 Jun 01 &   12.2&     22&       6&        1.63 $\pm$     0.46&                &                      &              &               &                           &   &low          \\
acis5938& 2005 Jun 03 &   12.0&     485&      185&      37.69$\pm$    1.87&     81$\pm$6.2&       12.0$\pm$5.8  &    1.89e-13&         3.0&             25.6&      2.041/17  &high\\
acis5939& 2005 Jun 06 &   12.0&     364&      181&      27.53$\pm$    1.61&     76$\pm$5.6&       8.3$\pm$4.2  &    1.62e-13&          2.6&             17.7&      0.857/13  &high\\
acis5940& 2005 Jun 09 &   12.1&     70&       20&       4.90 $\pm$     0.73&                &                      &              &               &                           &    &low         \\
acis5941& 2005 Jun 11 &   12.0&     429&      187&      32.16$\pm$    1.74&     91$\pm$5.8&       6.5$\pm$4.0  &    1.74e-13&         2.8&             10.6&      1.004/17 &high\\
acis5942& 2005 Jun 15 &   12.1&     405&      206&      30.30$\pm$    1.68&     70$\pm$5.4&       14.1$\pm$5.1  &    1.68e-13&         2.7&             42.5&      0.724/14 &high\\
acis5943& 2005 Jun 18 &   12.2&     525&      187&      41.30$\pm$    1.94&     91$\pm$5.4&       8.8$\pm$3.8   &    2.10e-13&         3.3&             16.0&      1.096/21 &high\\
acis5944& 2005 Jun 21 &   12.0&     356&      85&       28.71$\pm$    1.64&     96$\pm$9.8&       20.6$\pm$8.6  &    1.16e-13&         1.8&             20.3&      1.128/15 &high\\
acis5945& 2005 Jun 24 &   11.7&     415&      220&      32.04$\pm$    1.75&     65$\pm$4.4&       19.8$\pm$5.3  &    1.69e-13&         2.7&             97.4&       1.028/14 &high\\
acis5946& 2005 Jun 26 &   12.2&     287&      167&      20.75$\pm$    1.40&     70$\pm$7.1&       10.6$\pm$5.9  &    1.29e-13&         2.0&             22.5&      1.160/8 &high\\
acis5947& 2005 Jun 29 &   10.8&     40&       29&       2.81 $\pm$     0.62&                &                      &              &               &                           &      &low       \\
acis5948& 2005 Jul 03 &   12.2&     77&       51&       4.93 $\pm$     0.73&                &                      &              &               &                           &      &low       \\
acis5949& 2005 Jul 06 &   12.2&     54&       33&       3.36 $\pm$     0.62&                &                      &              &               &                           &      &low       \\
acis9805& 2007 Dec 21 &   5.2&      44&       28&       7.55 $\pm$     1.43&                &                      &              &               &                           &      &low       \\
acis9122& 2008 Feb 01 &   10.0&     55&       18&       5.40 $\pm$     0.87&                &                      &              &               &                           &      &low       \\
Total high&           &  149.3&   7085&     3584&      37.85$\pm$     2.53&   84$\pm$1.3   &       8.2$\pm$0.1    &     1.98e-13&         3.1  &              16.9        &     1.739/31 \\
Total low &           &   97.4&    384&      189&     3.61$\pm$     0.81&   82$\pm$8.3   &       5.6$\pm$11.3   &     2.03e-14&         0.3 &              1.4          &     1.069/12 \\
\hline
\end{tabular}
\end{center}
\end{sidewaystable}

\noindent{\bf Legends for Extended Data Tables}\\
{\bf Extended Data Table 1. Observations of M 81 ULS-1.}
The columns are:
(1) observation date,
(2) telescope,
(3) exposure time.
(4) spectral resolution. \\
{\bf Extended Data Table 2. Properties of H$_{\alpha}^-$, H$_{\alpha}^+(?)$ and H$_{\alpha}$.}
The center, FWHM and E.W. are in unit of \AA.
The number given in the bracket is the velocity in unit of the speed of light.
The power is in unit of 10$^{-16}$ erg s$^{-1}$ cm$^{-2}$.
All the errors refer to 68.3\% uncertainties. \\
{\bf Extended Data Table 3. Doppler boost factors for each observation.}
The columns are:
(1) observation date,
(2) wavelength of the blueshifted/redshifted $H_\alpha$,
(3) Doppler boosting factor,
(4) redshift. \\
{\bf Extended Data Table 4. {\it Chandra} Observations of M81 ULS-1.}
{\sc Notes:} Col. (1): ObsID. Col. (2): Obs Date. Col. (3): Ontime without
deadtime correction. Col. (4): Net counts in 0.1-8.0 keV. Col. (5): Counts in
the supersoft band, 0.1-0.5 keV. Col. (6): Count rate after vignetting
correction. Col. (7): Temperature for the blackbody fit to the spectrum in
0.3-8.0 keV. Col. (8): Neutral hydrogen column density. Col. (9): The 0.3-8.0
keV flux for the blackbody fit to the spectrum. (10): The 0.3-8.0 keV
luminosity. (11): The unabsorbed bolometric luminosity. Col. (12): Reduced
$\chi^2$ and degree of freedom for the spectral fit. Col. (13): Flags
indicating a high low state, seprated by Count Rate = 10 count ks$^{-1}$. \\

\includegraphics[width=150mm]{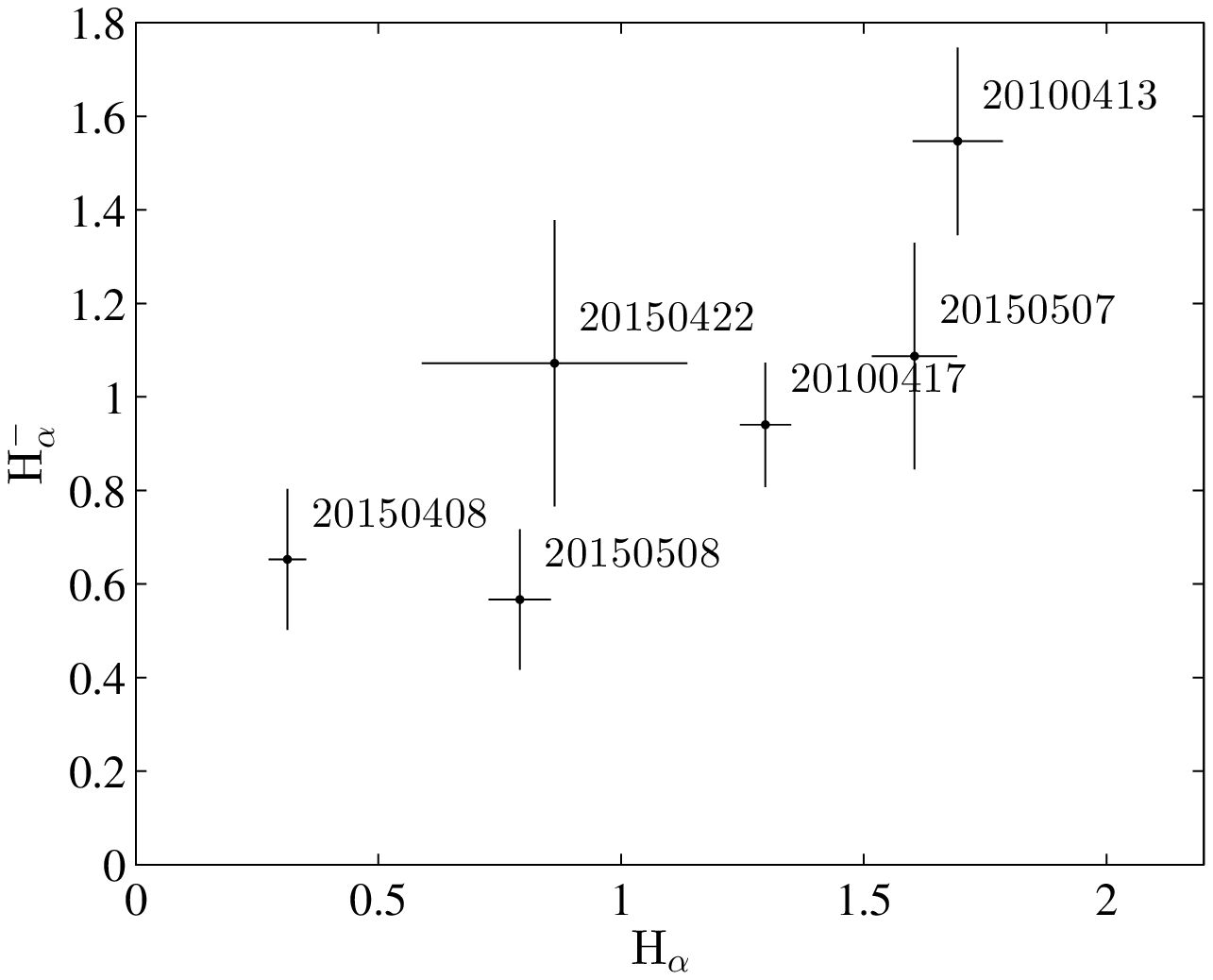} \\
\includegraphics[width=150mm]{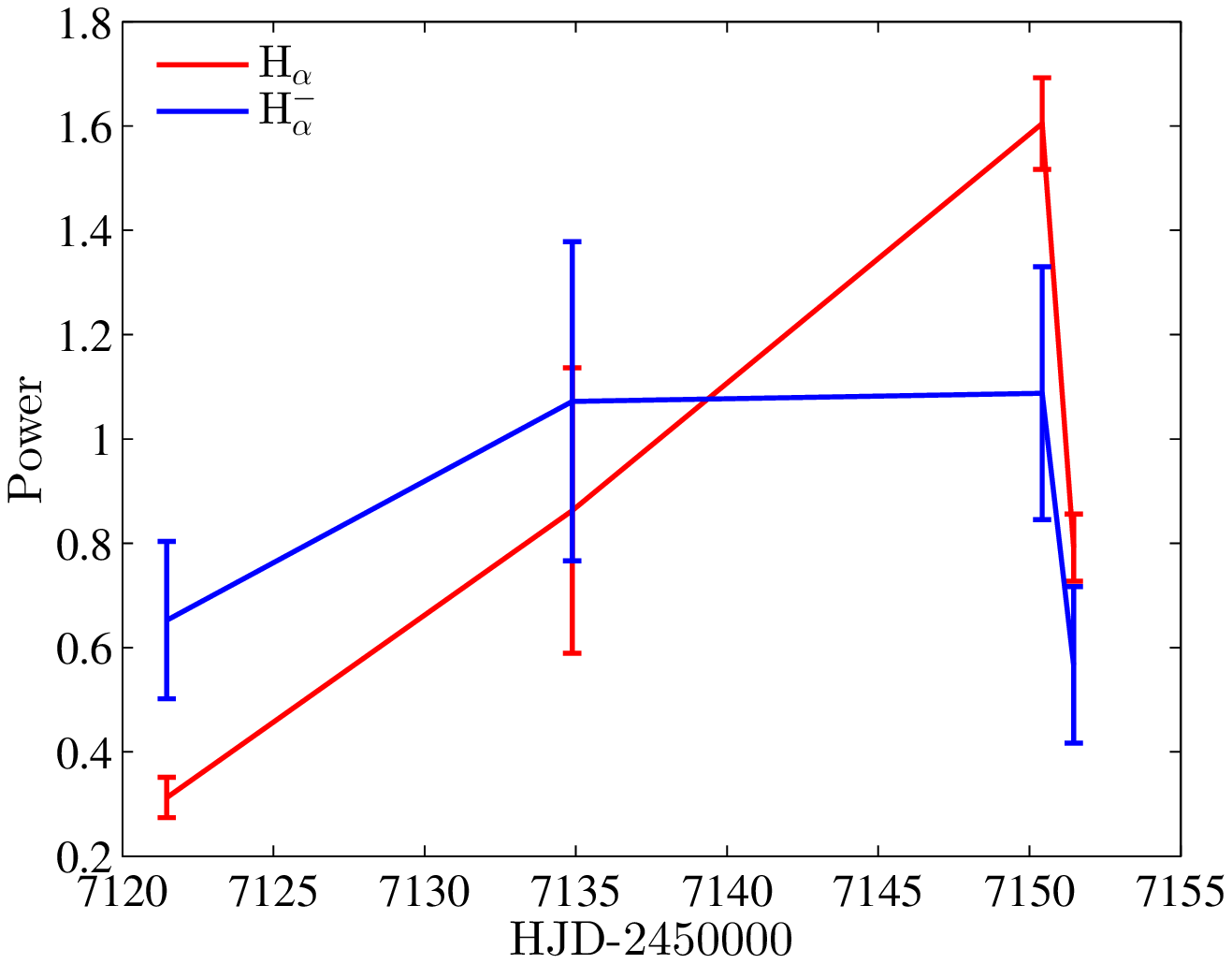} \\
\includegraphics[width=150mm]{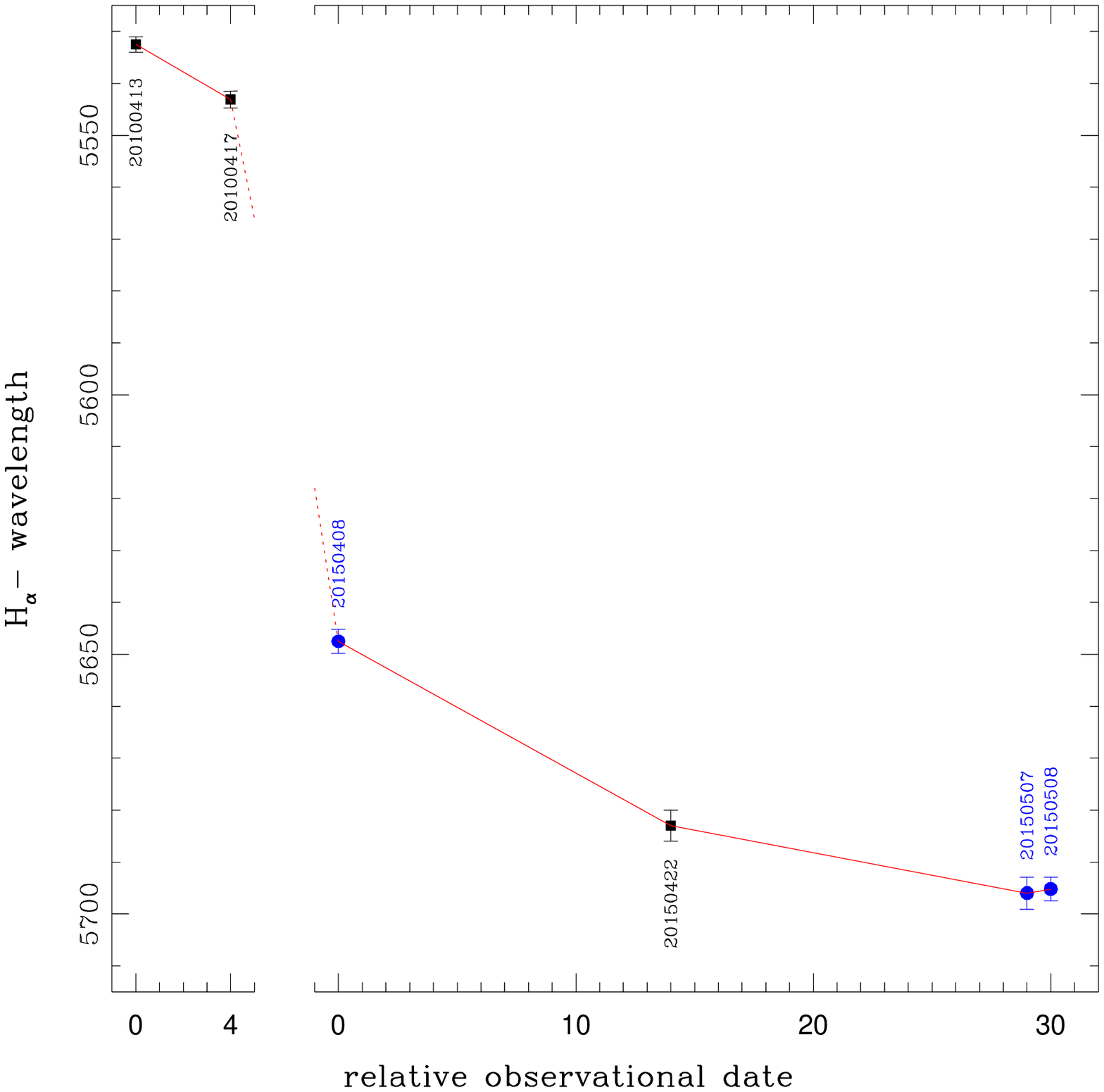} \\
\includegraphics[width=0.8\textwidth]{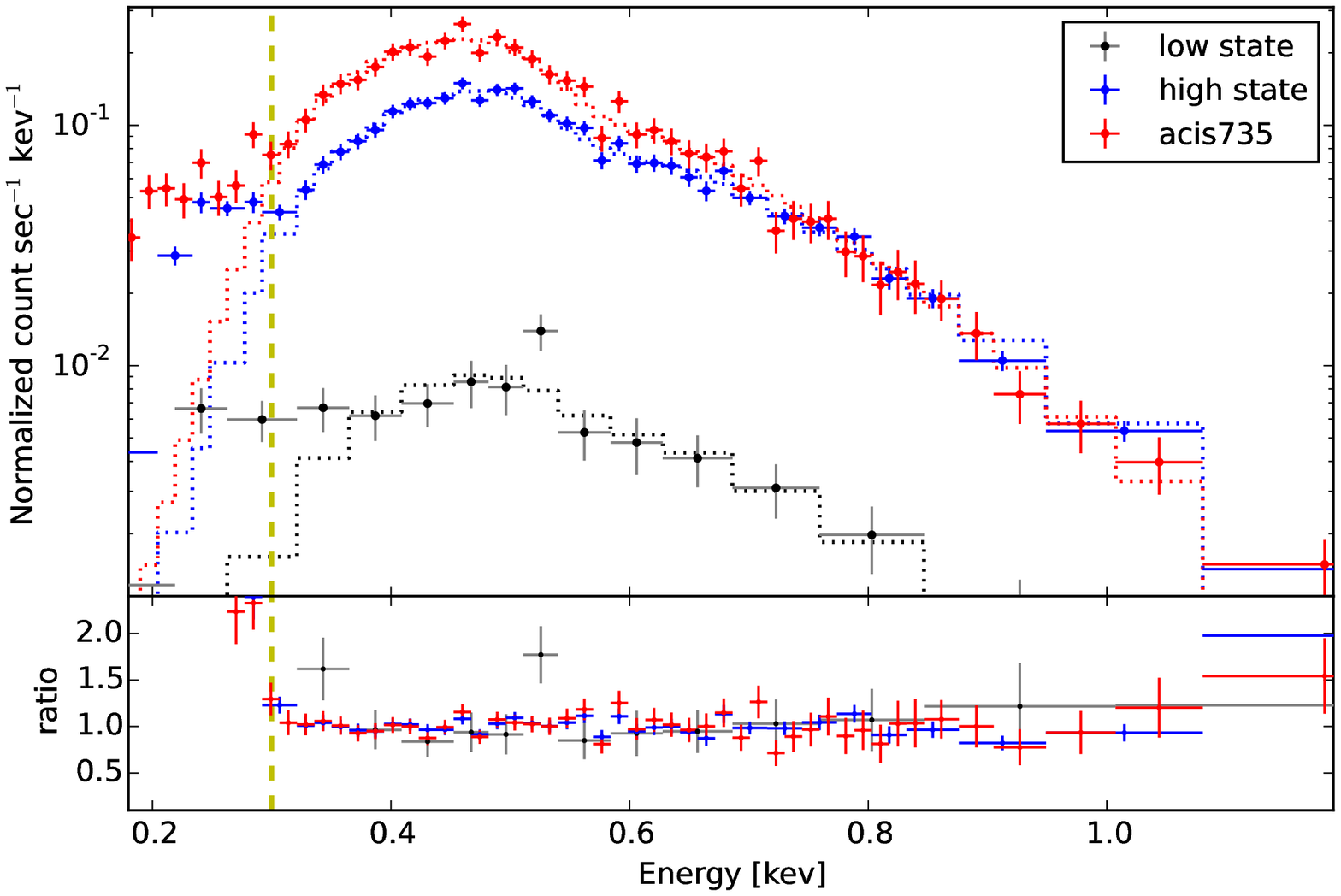}

\noindent{\bf Legends for Extended Data Figures}\\
{\bf Extended Data Figure 1.  The power of H$_{\alpha}^-$ vs. that of H$_{\alpha}$
         in unit of 10$^{-16}$ erg s$^{-1}$ cm$^{-2}$.} The error bars account for the 68.3\% uncertainties. \\
{\bf Extended Data Figure 2.  The variation of the power of emission lines
         in unit of 10$^{-16}$ erg s$^{-1}$ cm$^{-2}$.} The error bars account for the 68.3\% uncertainties.  \\
{\bf Extended Data Figure 3.  The center of H$_{\alpha}^-$ as a function of relative observational
         date.}  The dates of observations are marked in the figure. The error bars account for the 68.3\% uncertainties. \\
{\bf Extended Data Figure 4.  The M81 ULS-1 spectra from ObsID 735.} The spectra are combined high-state
observations, and combined low-state observations, shown with red, blue, and
black crosses, fitted by blackbody model in the 0.3-8.0 keV with red, blue, and
black dotted lines. The yellow dashed line indicates photon counts with energy
of 0.3 keV. The error bars account for the 68.3\% uncertainties. \\

\end{document}